\documentclass[a4paper]{appolb}
\usepackage{graphicx}
\usepackage{wrapfig}
\usepackage{amsmath}
\usepackage{amssymb}

\headauthor{A.~Sch\"alicke}
\headtitle{Polarised Positrons for the ILC}
\Preprinttrue

\begin{document}

\title{Polarised positrons for the ILC%
\thanks{Presented at the XXXI International Conference of Theoretical Physics, ``Matter to the Deepest'', Ustro\'n, Poland, September 5--11, 2007.}%
}
\author{Andreas Sch{\"a}licke
\address{DESY, Zeuthen, Germany}
}
\maketitle
\vspace{-5cm}
\hfill DESY-07-203\\[-6mm]
\vspace{5cm}
\begin{abstract}
For the planned International Linear Collider it is intended to
have both -- electron and positron -- beams polarised. This offers a
great benefit for many physics studies, but also provides a challenge
for the engineering of the machine. A polarised positron source that
meets the machine parameters is topic of current design studies and
prototype experiments.
\end{abstract}

\section{Introduction}
\sectionmark{Polarised positrons for the ILC}  
The International Linear Collider (ILC) is an electron-positron
collider, currently being in the design phase. The nominal
centre-of-mass energy is 500 GeV and the design luminosity
is $2 \times 10^{34} {\mbox{cm}}^{-2} {\mbox{s}}^{-1}$
\cite{elsen,RDR}, which makes the machine well adapted for precision
studies of the Standard Model (SM) and searches for new physics
\cite{maria}. Machine upgrades include the extension of the
energy range to 1 TeV, or a gamma-gamma and electron-gamma collider.

A key feature of the ILC 
is the possible polarisation of both beams (electrons and positrons),
which was shown to be of paramount
importance for many physics studies. But producing a high intensity
polarised positron beam proved to be a challenge.


The degree of polarisation for the electron beam is expected to be
at least 80\%. For the positron beam, an undulator based source will 
provide a degree of 30\% from the start, which will be improved to
60\% in a future upgrade of the machine. 


\section{Physics with polarised positrons}

The physics potential of the ILC utilising polarisation of electron
and positron beam has been investigated in great depth \cite{gudi}. Here
we like to give only a brief motivation and perhaps an answer to the
question: Why do we need polarised positrons? 

The first features that are usually addressed in this context are 
{\em higher effective luminosity and higher effective polarisation}:
When electron and positron annihilate into a vector particle 
(e.g.\ $\gamma$/$Z^o$), only two helicity combinations contribute to the total
cross section. Thus it is possible to write the polarisation dependent
cross section as
\begin{align}
    \sigma_{p_{e^-}p_{e^+}} 
  &= \left(1 - P_{e^+}P_{e^-}\right) \: \sigma_0 \: \left[1 \: - \:
             {P_{\rm eff}}\:A_{\rm LR} \right] \,,
  &
  P_{\rm eff}&=
  \frac{P_{e^-} - P_{e^+}}{1- P_{e^-} P_{e^+}}
\nonumber
\;,
\end{align}
where $\sigma_0=({\sigma_{\rm RL} + \sigma_{\rm LR}})/{4}$ denotes the unpolarised cross section, and 
$A_{\rm  LR}=({\sigma_{\rm LR}-\sigma_{\rm RL}})/ ({\sigma_{\rm LR}+\sigma_{\rm RL}})$  
gives the left-right asymmetry. The prefactor
$(1-P_{e^+}P_{e^-})$ is an effective luminosity increase. 
It can be seen that in case of no positron polarisation, the effective
polarisation $P_{\rm eff}$ reduces to the electron polarisation
$P_{e^-}$, on the other hand if positron polarisation $P_{e^+}$ is
present, the effective polarisation is increased (see also Table
\ref{tab:effpol}).
But the most important aspect  
is a reduced dependence on the polarimetry uncertainties.
In fact a measurement of the weak mixing angle $\sin\theta_{\rm eff}$
with a precision of ${\cal O}(10^{-5})$ is only possible by having both beams
polarised. 

\begin{table}
\tabcolsep3mm
\begin{center}
\begin{tabular}{|ll||c|c|c|}
\hline  
$P_{e^-}$, &  $P_{e^+}$ & $P_{\rm eff}$ & $(1 - P_{e^+}P_{e^-}) $ & $\Delta P_{\rm eff}/P_{\rm eff}$   \\ 
\hline
$0\%$,   & $0\%$  &  $0\%$     & 1.00 & 1.00 $\Delta P/P$\\
$-80\%$, & $0\%$  & $-80\%$    & 1.00 & 1.00 $\Delta P/P$\\
$-80\%$, &$+30\%$ & $-89\%$    & 1.24 & 0.57 $\Delta P/P$\\
$-80\%$, &$+60\%$ & $-95\%$    & 1.48 & 0.35 $\Delta P/P$\\
\hline
\end{tabular}
\end{center}
\caption{Effective luminosity, effective polarisation and its
  uncertainties for different beam polarisation setups. \label{tab:effpol}}
\end{table}

Another important application of positron polarisation is the {\em
  reduction of backgrounds}. For example the process $e^+e^-\to W^+W^-$, which
constitutes an important background to many beyond the SM physics
analyses, can be reduced by factor 10 when the corresponding electron (+80\%)
and positron (-60\%) polarisations are employed.

Positron polarisation can also help in the {\em separation between
  models of new physics}, by a direct probe of the spin in resonance
productions, which may occur in R-parity violating SUSY scenarios
(e.g.~ $e^+e^-\to \tilde{\nu}_{\tau} \to \mu^+ \mu^-$) or via heavy
gauge boson exchange ($e^+e^-\to Z^\prime \to \mu^+ \mu^-$) \cite{gudi,rparity}.

Furthermore, polarised positrons provide an unique possibility for
the understanding of non-standard couplings. For example, in the
super-\-sym\-met\-ric extension of the SM the separation of
$\tilde{e}^+_{\rm L} \tilde{e}^-_{\rm L}$  and $\tilde{e}^+_{\rm L}
\tilde{e}^-_{\rm R}$ is only possible with both beams being polarised \cite{gudi}. 
 
Other aspects of polarisation of both beams include {\em transverse
  polarisation}, which might prove helpful for the identification
signatures of extra dimensions in fermion production \cite{gudi,rizzo}. 

\section{The E166 experiment}

In the baseline design of the ILC \cite{RDR} polarised positrons are
produced from circularly polarised photons created in an helical
undulator hitting a thin Ti target. The spin of the photon is
transferred to the electron-positron pairs produced, resulting in
a net polarisation of the particles emerging from the target.  
The positrons are captured just behind the target in a dedicated
capture optics, i.e.\ an adiabatic matching device, and their degree
of polarisation has to be maintained until they reach the collision
point.  

A proof-of-principle experiment to demonstrate production of polarised
positrons in a manner suitable for implementation at the ILC has been
carried out at SLAC \cite{Alexander:2003fh}. A  
helical undulator of 2.54 mm period and 1-m length produced circularly
polarised photons of first harmonic endpoint energy of 8 MeV when
traversed by a 46.6 GeV electron beam. The polarised photons were
converted to polarised positrons in a 0.2-radiation-length tungsten
target. The polarisation of these positrons was measured at several
energies using a Compton transmission polarimeter. Yield and
polarisation of the photon beam was also continuously monitored.

\begin{figure}
\centerline{\includegraphics[width=0.65\columnwidth]{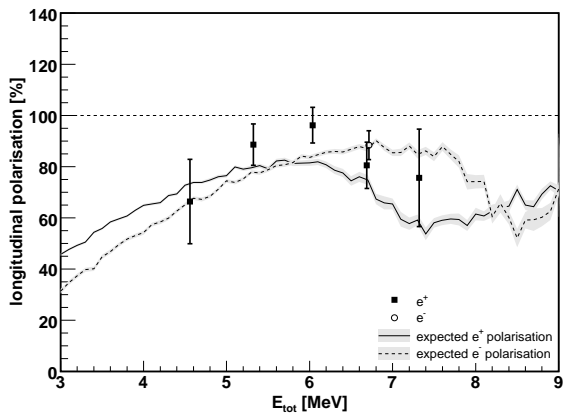}}
\caption{Preliminary E166 results of positron (and electron) beam polarisation.\label{fig:e166}}
\end{figure}
The experiment collected data during June and September 2005. The
measured asymmetries in the positron and photon polarimeters were
translated into polarisation, using the analysing power obtained in Geant 
simulations. 
Preliminary results \cite{e166proc,e166proc7} show good agreement
with the expectations. In Figure \ref{fig:e166} the measured degrees of
polarisation of positrons for 5 different energy points are
presented. At one energy point also the electron polarisation has been
measured. The results are compared with expectations obtained by Geant4
simulations using the new EM polarisation extension.


\section{Polarised Geant4}

Programs that can simulate the complex interaction patterns of
particles traversing matter are indispensable tools for the design and
optimisation of particle detectors. A major example of such programs
is Geant4 \cite{Agostinelli:2002hh,geant4}, which is widely used in
high energy physics, medicine and  
space science. Different parts of this tool kit can be combined to
optimally fulfil the users needs. A powerful geometry package allows
the creation of complex detector configurations. The physics
performance is based on a huge list of interaction processes. Tracking
of particles is possible in arbitrary electromagnetic fields.
However, polarisation has played only a minor role so far.

Starting with version 8.2 a new package of QED physics processes has
been added to the Geant4 framework, allowing studies of polarised
particles interactions with polarised media \cite{Dollan:2005nj,geant4lcws,geant4inprep}. 

The implementation of polarisation in the library in Geant4 follows
very closely the approach by \cite{McMaster:1961}. A {\em Stokes
  vector} is associated to each particle and used to track the 
polarisation from one interaction to another. 
Five new process classes for Bhabha/M{\o}ller scattering,
electron-positron annihilation, Compton scattering, pair creation, and 
bremsstrahlung with polarisation are now available for physics studies
with Geant4. The implementation has been carefully checked against
existing references, alternative codes, and dedicated analytic
calculations.  

\begin{figure}
\centerline{\includegraphics[height=4.5cm,width=0.45\columnwidth]{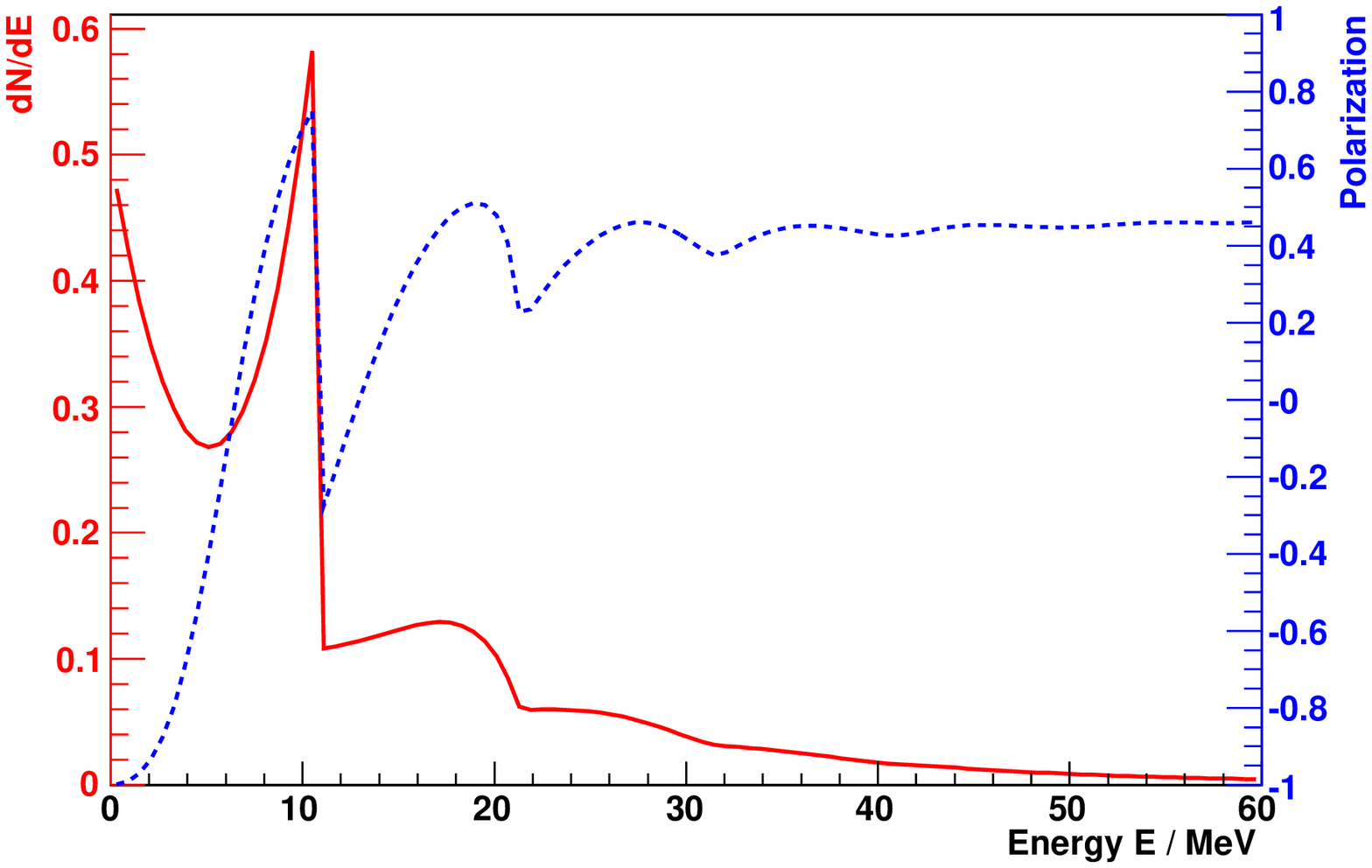}
            \includegraphics[height=4.5cm,width=0.45\columnwidth]{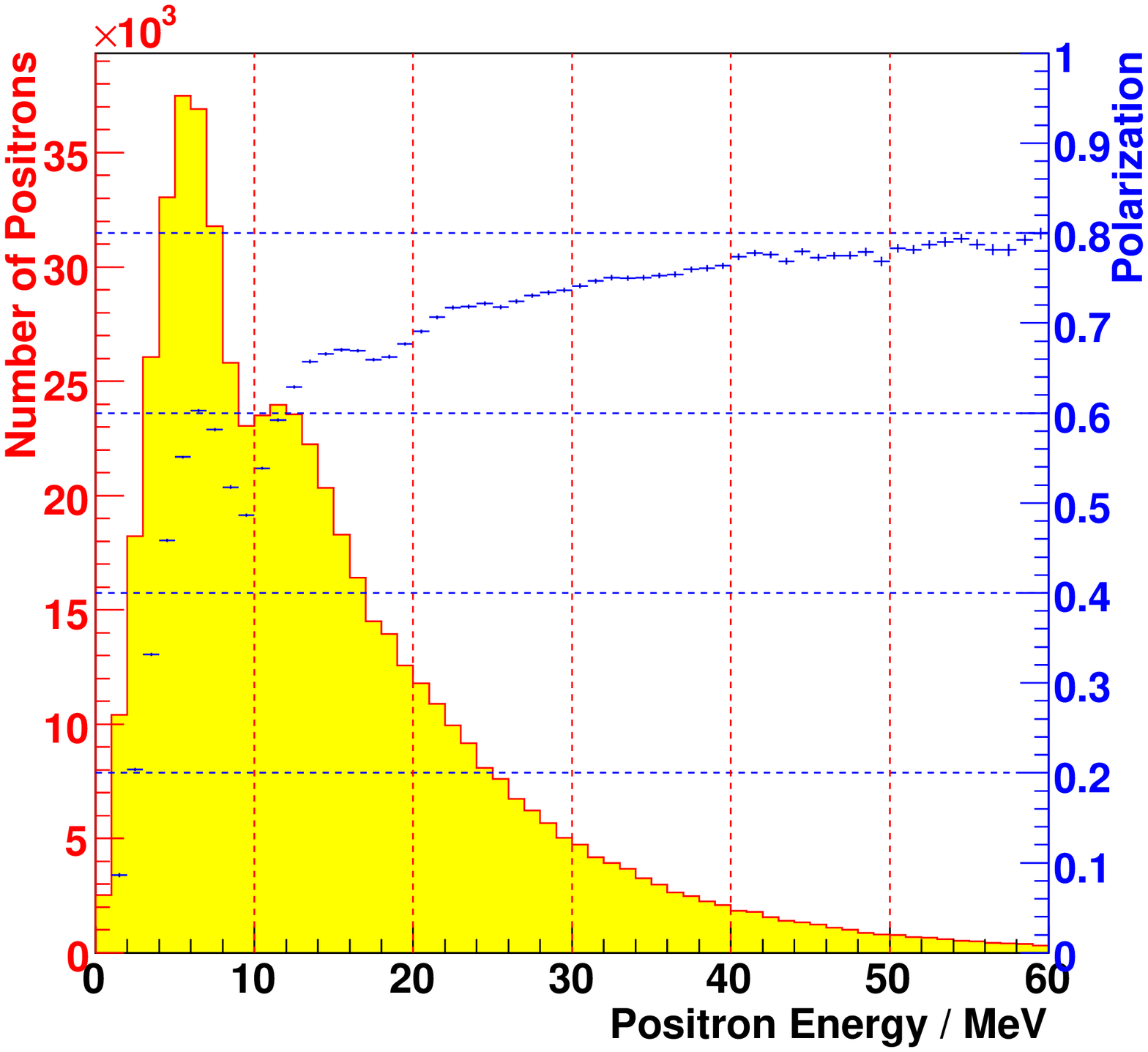}}

\caption{Left: Energy (solid) and polarisation (dashed) of photons created
  in a helical undulator.  Right: Energy (histogram) and polarisation
  (dashed line) of positrons after the production target of  thickness
  $d=0.4 X_0$ 
}
\label{fig:ilcPPS}
\end{figure}

Applications include
design and optimisation of a polarised positron source and beam
polarimetry for a future linear collider facility. Figure
\ref{fig:ilcPPS} shows the energy spectrum and degree of polarisation
for photons created in helical undulator with strength $K=1$ and
period $\lambda=1\,$cm, as well as the corresponding energy and
polarisation spectra for the produced positrons.

\section{Summary}
Having not only the electron beam but also the positron beam polarised 
is highly advantageous for most physics studies, vital for many
precision measurements, and essential for 
model analyses.
The E166 experiment successfully demonstrated the undulator based
production scheme for polarised positrons. A newly developed extension
to Geant4 now allows to simulate the interactions of polarised
particles with (polarised) matter, which are already used in many ongoing
R\&D around the polarised positron source.

\section*{Acknowledgements}
The author likes to thank all colleagues involved in R\&D of the
ILC positron source. Particular thanks goes to my collaborators in
the E166 experiment, the LEPOL Collaboration, and the Geant4
Collaboration.

\end{document}